\documentclass[11pt]{article}
\begin{document}

\begin{center}{\Large Computation of the Gradient and the Hessian of the Log-likelihood 
of the State-space Model by the Kalman Filter}\\

\vspace{5mm}
{\large Genshiro Kitagawa}\\
Mathematics and Informatics Center, The University of Tokyo\\
and Meiji Institute for Advanced Study of Mathematical Sciences, Meiji University

\vspace{3mm}
{\today}

\end{center}

\noindent{\bf Abstract}

The mazimum likelihood estimates of an ARMA model can be obtained by the Kalman filter
based on the state-space representation of the model.
This paper presents an algorithm for computing gradient of the log-likelihood
by an extending the Kalman filter without resorting to the numerical difference.
Three examples of seasona ledjustment model and ARMA model are presented to
exemplified the specification of structural matrices and initaial matrices.
An extension of the algorithm to compute the Hessian matrix is also shown. \\


\section{Introduction: The Maximum Likelihood Estimation of a State-Space Model}

We consider a linear Gaussian state-space model 
\begin{eqnarray}
 x_n &=& F_n(\theta) x_{n-1} + G_n(\theta) v_{n} \label{Eq-SSM-1}\\
 y_n &=& H_n(\theta) x_n + w_n,  \label{Eq-SSM-2}
\end{eqnarray}
where $y_n$ is a one-dimensional time series, $x_n$ is an $m$-dimensional state vector, $v_n$ is a $k$-dimesional Gaussian white noise, $v_n \sim N(0,Q_n(\theta ))$, and $w_n$ is one-dimensional white noise, $w_n \sim N(0,R_n(\theta ))$.
$F_n(\theta )$, $G_n(\theta )$ and $H_n(\theta )$ are $m\times m$ matrix, $m\times k$ matrix and $m$ vector. respectively.
$\theta$ is the $p$-dimensional parameter vector of the state-space model such as the variances of the noise
inputs and unknown coefficients in the matrices $F_n(\theta )$, $G_n(\theta )$, $H_n(\theta )$, $Q_n(\theta )$ and $R_n(\theta )$.
For simplicity of the notation, hereafter, the parameter $\theta$ and the suffix $n$ will be omitted.

Various models used in time series analysis can be treated uniformly within the state-space model framework. 
Further, many problems of time series analysis, such as prediction, signal extraction, decompositoon, parameter estimation and interpolation, can be formulated as the state estimation of a state-space model.

Given the time series $Y_N\equiv \{y_1,\ldots ,y_N\} $ and the state-space model (\ref{Eq-SSM-1})nd  
and (\ref{Eq-SSM-2}), the one-step-ahead predictor $x_{n|n-1}$ and the filter $x_{n|n}$ and their variance covariance 
matrices $V_{n|n-1}$ and $V_{n|n}$ are obtained by the Kalman filter (Anderson and Moore (2012) and Kitagawa (2020)):

One-step-ahead prediction
\begin{eqnarray}
 x_{n|n-1} &=& F x_{n-1|n-1} \nonumber \\
 V_{n|n-1} &=& F V_{n-1|n-1} F^T + G Q_{n} G^T \label{Eq-3-2}
\end{eqnarray}
\indent
Filter 
\begin{eqnarray}
 K_n &=& V_{n|n-1}H^T (H V_{n|n-1} H^T + R)^{-1} \nonumber \\
 x_{n|n} &=& x_{n|n-1} + K_n (y_n -H x_{n|n-1}) \label{Eq-3-3} \\
 V_{n|n} &=& (I -K_n H)V_{n|n-1}. \nonumber
\end{eqnarray}

Given the data $Y_N$, the likelihood of the time series model is defined by 
\begin{eqnarray}
 L(\theta ) &=& p( Y_N|\theta ) 
  =  \prod_{n=1}^N g_n(y_n|Y_{n-1},\theta ),
\end{eqnarray}
where $g_n(y_n|Y_{n-1},\theta )$ is the conditional distribution of $y_n$ given the observation $Y_{n-1}$ and is a normal distribution given by
\begin{eqnarray}
g_n(y_n|Y_{n-1},\theta ) 
  = \frac{1}{\sqrt{2\pi r_n}}\exp\left\{ -\frac{\varepsilon_{n}^2}{2r_n} \right\}, 
\label{eq_distrribution_g}
\end{eqnarray}
where $\varepsilon_n$ and $r_n$ are the one-step-ahead prediction error and
its variance defined by
\begin{eqnarray}
\varepsilon_n &=& y_n - Hx_{n|n-1} \nonumber \\
r_n &=&  H_n V_{n|n-1} H_n^T + R  
\end{eqnarray}

Therefore, the log-likelihood of the state-space model is obtained as
\begin{eqnarray}
 \ell (\theta ) = \log L(\theta ) &=& \sum_{n=1}^N \log g_n(y_n|Y_{n-1},\theta ) \nonumber \\
  &=& -\frac{1}{2} \biggl\{ N \log 2\pi + \sum_{n=1}^N \log r_{n}
 + \sum_{n=1}^N \frac{\varepsilon_{n}^2}{r_n} \biggr\}. \label{Eq_log-lk}
\end{eqnarray}

The maximum likelihood estimates of the parameters of the state-space model 
can be obtained by maximizing the log-likelihood function.
In general, since the log-likelihood function is mostly nonlinear, the maximum likelihood estimates  is obtained by using a numerical optimization algorithm based on the quasi-Newton method. 
According to this method, using the value  $\ell( \theta)$ of the log-likelihood and the first derivative (gradient) $\partial \ell/\partial \theta$ for a given parameter \( \theta \), the maximizer of \( \ell( \theta) \) is automatically estimated by repeating 
\begin{equation}
 \theta_k = \theta_{k-1} + \lambda_k B_{k-1}^{-1} \frac{\partial \ell}{\partial \theta},
\end{equation}
where $\theta_0$ is an initial estimate of the parameter. 
The step width \( \lambda_k \) is automatically determined and the inverse matrix \( H_{k-1}^{-1} \) of the Hessian matrix is obtained recursively by the DFP or BFGS algorithms (Fletcher (2013)).

Here, the gradient of the log-likelihood function is usually approximated by numerical difference,
such as 
\begin{eqnarray}
\frac{\partial \ell (\theta)}{\partial \theta_j} \approx \frac{\ell (\theta_j +\Delta \theta_j) -  \ell (\theta_j  \Delta \theta_j)}{2\Delta\theta_j},
\end{eqnarray}
where $\Delta\theta_j$ is defined by $C |\theta_j|$, for some small $C$ such as 0.00001.
The numerical difference usually yields reasonable approximation to the gradient of the 
log-likelihood.
However, since it requires $2p$ times of log-likelihood evaluations, the amount of computation
becomes considerable if the dimension of the parameters is large.
Further, if the the maximum likelihood estimates lie very close to the boundary of
addmissible domain, which sometimes occure in regularization problems,
it becomes difficlt to obtain the approximation to the gradient of the log-likelihood
by  the numerical difference.

Analytic derivative of the log-likelihood of time series models were considered by many authors. 
For example,  Kohn and Ansley (1985) gave method for computing likelihood and its derivatives for an ARMA model.
Zadrozny (1989) derived analytic derivatives for estimation of linear dynamic models.
Kulikova (2009) presented square-root algorithm for the likelihood gradient evaluation to avoid numerical instatbility of the recursive algorithm for log-likelihood computation.
In this paper,  the gradient and Hessian of the log-likelihood of linear state-space model are given. 
Details of the implementation of the algorithm for standard seasonal adjustment model, seasonal adjustment model with stationary AR component and ARMA model are given.
For each implementation, comparison with a numerical difference method is shown.

In section 2, we consider to obtain the gradient of the log-likelihood
by extending the Kalman filter algorith.
Extension of the algorithm for computing the Hessian of the log-likelihood is shown in section 3.
Application of the method is exemplified with the three models, i.e., the standard seasonal adjustment model, the seasonal adjustment model
with autoregressive component, and ARMA (autoregressive moving average model) are shown in section 4.

\section{The Gradient and the Hessian of the log-likelihood }
\subsection{The gradient of the log-likelihood}

From (\ref{Eq_log-lk}), the gradient of the log-likelihood is obtained by
\begin{eqnarray}
\frac{\partial\ell (\theta )}{\partial \theta} 
&=& - \frac{1}{2}\sum_{n=1}^N \left\{ \frac{1}{r_n}\frac{\partial r_n}{\partial\theta}
    + 2 \frac{\varepsilon_n}{r_n}  \frac{\partial \varepsilon_n}{\partial\theta}
    -  \frac{\varepsilon_n^2}{r_n^2}\frac{\partial r_n}{\partial\theta}
   \right\},  \label{Eq_gradient_ell}
\end{eqnarray}
where, from (\ref{eq_distrribution_g}), the derivatives of the one-step-ahead predition $\varepsilon_n$ and the
one-step-ahead prediction error variance $r_n$ are obtained by
\begin{eqnarray}
\frac{\partial \varepsilon_n}{\partial\theta} &=& -H \frac{\partial x_{n|n-1}}{\partial\theta}
   - \frac{\partial H}{\partial\theta}x_{n|n-1} \nonumber \\
\frac{\partial r_n}{\partial\theta} &=& H \frac{\partial V_{n|n-1}}{\partial\theta}H^T
   + \frac{\partial H}{\partial\theta}V_{n|n-1}H^T 
   + H V_{n|n-1} \frac{\partial H}{\partial\theta}^T
   + \frac{\partial R}{\partial\theta} . \label{Eq_gradient_observation_error}
\end{eqnarray}
To evaluate these quantity, we need the derivative of the one-step-ahead predictor
of the state $\displaystyle\frac{\partial x_{n|n-1}}{\partial\theta}$ and its variance covariance
matrix $\displaystyle\frac{\partial V_{n|n-1}}{\partial\theta}$ which can be obtained recursively
in parallel to the Kalman filter algorithm:\\

\quad [One-step-ahead-prediction] \\
\begin{eqnarray}
\frac{\partial x_{n|n-1}}{\partial\theta} &=& F \frac{\partial x_{n-1|n-1}}{\partial\theta}
   + \frac{\partial F}{\partial\theta} x_{n-1|n-1}
\nonumber \\
\frac{\partial V_{n|n-1}}{\partial\theta} &=& F \frac{\partial V_{n-1|n-1}}{\partial\theta}F^T 
           + \frac{\partial F}{\partial\theta}V_{n-1|n-1}F^T 
           + F V_{n-1|n-1} \frac{\partial F}{\partial\theta}^T \nonumber \\
         &&  + G\frac{\partial Q}{\partial\theta}G^T 
           + \frac{\partial G}{\partial\theta}Q G^T 
           + G Q \frac{\partial G}{\partial\theta}^T .  \label{Eq_gradient-filter-P}
\end{eqnarray}

\quad [Filter] \\
\begin{eqnarray}
\frac{\partial K_n}{\partial\theta} 
        &=& \left(\frac{\partial V_{n|n-1}}{\partial\theta}H^T
               +   V_{n|n-1}\frac{\partial H}{\partial\theta}^T\right) r_n^{-1} 
          - V_{n|n-1}H^Tr_n^{-2 }
          \frac{\partial r_n}{\partial\theta}    \nonumber \\
\frac{\partial x_{n|n}}{\partial\theta} &=&  
        \frac{\partial x_{n|n-1}}{\partial\theta}
       + K_n \frac{\partial \varepsilon_n}{\partial\theta}
       + \frac{\partial K_n}{\partial\theta} \varepsilon_n   \nonumber \\
\frac{\partial V_{n|n}}{\partial\theta} &=& 
         \frac{\partial V_{n|n-1}}{\partial\theta}
       - \frac{\partial K_n}{\partial\theta} H V_{n|n-1}
       - K_n \frac{\partial H}{\partial\theta} V_{n|n-1}
       - K_n H \frac{\partial V_{n|n-1}}{\partial\theta}. \label{Eq_gradient-filter-F}
\end{eqnarray}

\subsection{Hessian of the Log-likelihood of the State-space Model}

The Hessian (the second derivative) of the log-likelihood is also obtained
by a recursive formula, since, from (\ref{Eq_gradient_ell}), it is given as
%
%
\begin{eqnarray}
\frac{\partial^2\ell (\theta )}{\partial \theta\partial \theta^T}
&=& - \frac{1}{2}\sum_{n=1}^N \left\{ 
\frac{1}{r_n}\left(\frac{\partial^2 r_n}{\partial\theta\partial\theta^T} + \frac{\partial \varepsilon_n}{\partial\theta}\frac{\partial \varepsilon_n}{\partial\theta^T} \right) 
- \frac{1}{r_n^2} \left( 
\frac{\partial r_n}{\partial\theta}\frac{\partial r_n}{\partial\theta^T}
 - \varepsilon_n \frac{\partial r_n}{\partial\theta} \frac{\partial \varepsilon_n}{\partial\theta^T} 
\right.\right. \nonumber\\
&& \left.\left.
 - \varepsilon_n \frac{\partial \varepsilon_n}{\partial\theta} \frac{\partial r_n}{\partial\theta^T} 
 +  \varepsilon_n^2\frac{\partial^2 \varepsilon_n}{\partial\theta \partial\theta^T} 
 + \frac{\varepsilon_n^2}{2} \frac{\partial^2 r_n}{\partial\theta\partial\theta^T}
\right)  
- \frac{\varepsilon_n^2}{r_n^3} \frac{\partial r_n}{\partial\theta}\frac{\partial r_n}{\partial\theta^T}
\right\},
\end{eqnarray}
where, from (\ref{Eq_gradient_observation_error}), $ \displaystyle\frac{\partial^2 \varepsilon_n}{\partial\theta\partial\theta^T} $
and $ \displaystyle\frac{\partial^2 r_n}{\partial\theta \partial\theta^T} $ are
obtained by
\begin{eqnarray}
\frac{\partial^2 \varepsilon_n}{\partial\theta\partial\theta^T} 
&=&  -  2\frac{\partial H}{\partial\theta}\frac{\partial x_{n|n-1}}{\partial\theta^T}
-  H\frac{\partial^2 x_{n|n-1}}{\partial\theta\partial\theta^T}
-  \frac{\partial^2 H}{\partial\theta \partial\theta^T}x_{n|n-1} 
 \nonumber \\
\frac{\partial^2 r_n}{\partial\theta \partial\theta^T} 
   &=& 2\frac{\partial H}{\partial\theta}\frac{\partial V_{n|n-1}}{\partial\theta^T}H^T
     + H \frac{\partial^2 V_{n|n-1}}{\partial\theta \partial\theta^T}H^T
     + 2H \frac{\partial V_{n|n-1}}{\partial\theta}\frac{\partial H}{\partial\theta^T} \\
   &&+ \frac{\partial^2 H}{\partial\theta\partial\theta^T}V_{n|n-1}H^T 
     + 2\frac{\partial H}{\partial\theta}V_{n|n-1}\frac{\partial H^T}{\partial\theta^T} 
     + H V_{n|n-1} \frac{\partial^2 H}{\partial\theta \partial\theta^T}
     + \frac{\partial^2 R}{\partial\theta\partial\theta^T}. \nonumber 
\end{eqnarray}

Therefore, to evaluate the Hessian, the following computation should be performed
along with the recursive formula for thelog-likelihood and the 
gradient of the log-likelihood.%

\begin{eqnarray}
\frac{\partial^2 x_{n|n-1}}{\partial\theta \partial\theta^T}
   &=& 2\frac{\partial F}{\partial\theta} \frac{\partial x_{n-1|n-1}}{\partial\theta^T}
     + F \frac{\partial^2 x_{n-1|n-1}}{\partial\theta \partial\theta^T}
     + \frac{\partial^2 F}{\partial\theta \partial\theta^T} x_{n-1|n-1}
\nonumber \\
\frac{\partial^2 V_{n|n-1}}{\partial\theta \partial\theta^T}
         &=& 2\frac{\partial F}{\partial\theta} \frac{\partial V_{n-1|n-1}}{\partial\theta^T}F^T  
          + F \frac{\partial^2 V_{n-1|n-1}}{\partial\theta \partial\theta^T}F^T  
          + 2F \frac{\partial V_{n-1|n-1}}{\partial\theta}\frac{\partial F^T}{\partial\theta^T}  \nonumber \\
       && + \frac{\partial^2 F}{\partial\theta \partial\theta^T}V_{n|n-1}F^T 
          + 2\frac{\partial F}{\partial\theta}V_{n|n-1}\frac{\partial F^T}{\partial\theta^T} 
          + F V_{n|n-1} \frac{\partial^2 F^T}{\partial\theta \partial\theta^T} \nonumber \\
       && + 2\frac{\partial G}{\partial\theta} \frac{\partial Q}{\partial\theta^T}G^T 
          + G\frac{\partial^2 Q}{\partial\theta \partial\theta^T}G^T 
          + 2G\frac{\partial Q}{\partial\theta}\frac{\partial G^T}{\partial\theta^T} \nonumber \\
       && + \frac{\partial^2 G}{\partial\theta\partial\theta^T}Q G^T 
          + 2\frac{\partial G}{\partial\theta}Q \frac{\partial G^T}{\partial\theta^T} 
          + G Q \frac{\partial^2 G^T}{\partial\theta \partial\theta^T} \nonumber \\
\frac{\partial^2 K_n}{\partial\theta \partial\theta^T} 
        &=& \left(\frac{\partial^2 V_{n|n-1}}{\partial\theta \partial\theta^T}H^T  
             + 2\frac{\partial V_{n|n-1}}{\partial\theta}\frac{\partial H}{\partial\theta^T}
             + V_{n|n-1}\frac{\partial^2 H}{\partial\theta \partial\theta^T}
           \right) r_n^{-1} \nonumber \\
       && - 2\left(\frac{\partial V_{n|n-1}}{\partial\theta}H^T  
             + V_{n|n-1}\frac{\partial H}{\partial\theta}^T\right) r_n^{-2}\frac{\partial r_n}{\partial\theta^T}  \nonumber \\
       && + 2V_{n|n-1}H^T r_n^{-3}  
            \frac{\partial r_n}{\partial\theta} \frac{\partial r_n}{\partial\theta^T} 
        - V_{n|n-1}H^T r_n^{-2}  \frac{\partial^2 r_n}{\partial\theta \partial\theta^T}    \\
\frac{\partial^2 x_{n|n}}{\partial\theta \partial\theta^T} &=&  
        \frac{\partial^2 x_{n|n-1}}{\partial\theta \partial\theta^T}
       + 2 \frac{\partial K_n}{\partial\theta} \frac{\partial \varepsilon_n}{\partial\theta^T}
       + K_n \frac{\partial^2 \varepsilon_n}{\partial\theta \partial\theta^T} 
       + \frac{\partial^2 K_n}{\partial\theta \partial\theta^T} \varepsilon_n
\nonumber \\
\frac{\partial^2 V_{n|n}}{\partial\theta \partial\theta^T} &=& 
         \frac{\partial^2 V_{n|n-1}}{\partial\theta \partial\theta^T}
       - \frac{\partial^2 K_n}{\partial\theta \partial\theta^T} H V_{n|n-1}
       - 2\frac{\partial K_n}{\partial\theta} \frac{\partial H}{\partial\theta^T} V_{n|n-1}
       - 2\frac{\partial K_n}{\partial\theta} H \frac{\partial V_{n|n-1}}{\partial\theta^T} \nonumber \\
    &&
       - K_n \frac{\partial^2 H}{\partial\theta \partial\theta^T} V_{n|n-1}
       - 2K_n \frac{\partial H}{\partial\theta} \frac{\partial V_{n|n-1}}{\partial\theta^T} 
       - K_n H \frac{\partial^2 V_{n|n-1}}{\partial\theta \partial\theta^T}. \nonumber
\end{eqnarray}

\section{Examples}
In order to impliment the grafient filter, it is necessary to to specify the derivatives
of $F$, $G$, $H$, $Q$ and $R$ along with the original state-space model.
In this section, we shall consider three typical cases.
The first example is the standard seasonal adjeustment model, for which three matrics (or
vector), $F$, $G$ and $H$ do not contain unknown parameters and thus the derivatives
of these matrics becomes 0. 
This makes the algorithm for the gradient of the log-likelihood considerablly simple.
The second example is the seasonal adjustment model with AR component.
For this model,  the matrix $F$ depends on the unknown AR coefficients, 
but the derivative of $F$ is very simple and very sparse.
On the other hand, if we use a nonlinear transformation of the parameters 
in estimating the AR coefficients, to ensure the stationarity condition,
it is necessary to consider the effect of the transformation.
The third example is the ARMA model. 
Since the variance covariance matrix of the initial state vector
is complex functions of the AR and MA parameter, it is rather raborious
work to detemine the initial matrix for the algorithm for the gradient
of the log-likelihood.

\subsection{The standard seasonal adjustment model}

This is a typical example of the case where only the noise covariances $Q$ and $R$ 
depend on the unknown parameter $\theta$.
Consider a standard seasonal adjustment model
\begin{eqnarray}
y_n = T_n + S_n + w_n,
\end{eqnarray}
where $T_n$ and $S_n$ are the trend component and the seasonal component
that typically follow the following model
\begin{eqnarray}
&&T_n = 2T_{n-1} - T_{n-2} + u_n, \nonumber \\
&&S_n =-(S_{n-1}+\cdots +S_{n-p+1}) + v_n.
\end{eqnarray}
$u_n$, $v_n$ and $w_n$ are assumed to be Gaussian white noise with
variances $\tau_1^2$, $\tau_2^2$ and $\sigma^2$, respectively (Kitagawa and Gersch (1984,1996) and 
Kitagawa (2020)).

This seasonal adjustment model with two component models can be expressed in
state-space model form as
\begin{eqnarray}
x_n &=& F x_{n-1} + G v_n \nonumber \\
y_n &=& H x_n + w_n
\end{eqnarray}
with $v_n \sim N(0,Q)$ and $w_n \sim N(0,R)$ and the state vector $x_n$ and 
the matrices $F$, $G$, $H$, $Q$ and $R$ are defined by
{\setlength{\arraycolsep}{1mm}
\begin{eqnarray}
x_n = \left[ \begin{array}{c} 
             T_n     \\
             T_{n-1} \\
             S_n     \\
             S_{n-1} \\
            \vdots\\ 
             S_{n-p+2} \end{array}\right],\quad
F &=& \left[ \begin{array}{cccccc} 
             2 &-1 &   &   &   &   \\
             1 & 1 &   &   &   &   \\
               &   &-1 &-1 &\cdots&-1\\
               &   & 1 &   &   &   \\
               &   &   &\ddots&&   \\ 
               &   &   &   & 1 &    \end{array}\right],\quad
G = \left[ \begin{array}{cc} 
             1 & 0   \\
             0 & 0  \\
             0 & 1  \\
             0 & 0  \\
          \vdots&\vdots\\ 
             0 & 0   \end{array}\right] \\
H &=& [\begin{array}{cccccc} 1&0&1&0&\cdots &0 \end{array}] \nonumber \\
Q &=& \left[ \begin{array}{cc} 
             \tau_1^2 & 0   \\
               0      & \tau_2^2 \end{array}\right], \quad
R = \sigma^2.
\end{eqnarray}  
}

In this case, the parameter is $\theta =(\tau_1^2,\tau_2^2,\sigma^2)^T$,
and the $F$, $G$ and $H$ do not depend on the parameter.
Further, all of $F$, $G$, $H$, $Q$ and $R$ are time-invariant and
do not depend on time $n$.

In actual likelihood maximization, since there are positivity 
constrains, $\tau_1^2 > 0$, $\tau_2^2 >0$ and $\sigma^2 >0$,
it is frequently used a log-transformation,
\begin{eqnarray}
 \theta_1 = \log (\tau_1^2), \quad
 \theta_2 = \log (\tau_2^2), \quad
 \theta_3 = \log (\sigma^2) .
\end{eqnarray}

In this case, 
\begin{eqnarray}
 && \frac{\partial Q}{\partial\theta_1} =\left[ \begin{array}{cc} 
             \tau_1^2 & 0   \\
               0      & 0 \end{array}\right], \quad
  \frac{\partial Q}{\partial\theta_2} =\left[ \begin{array}{cc} 
               0 & 0   \\
               0 & \tau_2^2  \end{array}\right], \quad
  \frac{\partial Q}{\partial\theta_3} =\left[ \begin{array}{cc} 
               0 & 0   \\
               0 & 0  \end{array}\right], \quad \\
&&\frac{\partial R}{\partial\theta_1} = 0, \quad
  \frac{\partial R}{\partial\theta_2} = 0, \quad
  \frac{\partial R}{\partial\theta_3} = \sigma^2. \quad
\end{eqnarray}
 
Since $F$, $G$ and $H$ do not depend on $\theta$ and $\displaystyle\frac{\partial F}{\partial\theta}=0$,
$\displaystyle\frac{\partial G}{\partial\theta}=0$ and $\displaystyle\frac{\partial H}{\partial\theta}=0$ hold,
the recursive algorithm for gradient of the log-likelihood shown in (\ref{Eq_gradient-filter-P})
and (\ref{Eq_gradient-filter-F})become simple as follows:
\begin{eqnarray}
\frac{\partial \varepsilon_n}{\partial\theta} &=& -H \frac{\partial x_{n|n-1}}{\partial\theta}
    \nonumber \\
\frac{\partial r_n}{\partial\theta} &=& H \frac{\partial V_{n|n-1}}{\partial\theta}H^T
    + \frac{\partial R}{\partial\theta}     \nonumber \\%
%
%
\frac{\partial x_{n|n-1}}{\partial\theta} &=& F \frac{\partial x_{n-1|n-1}}{\partial\theta}
\nonumber \\
\frac{\partial V_{n|n-1}}{\partial\theta} &=& F \frac{\partial V_{n-1|n-1}}{\partial\theta}F^T 
           + G\frac{\partial Q}{\partial\theta}G^T \nonumber \\
\frac{\partial K_n}{\partial\theta} 
        &=& \frac{\partial V_{n|n-1}}{\partial\theta}H^T r_n^{-1}  
      - V_{n|n-1}H^Tr_n^{-2} \frac{\partial r_n}{\partial\theta}  \nonumber  \\
\frac{\partial x_{n|n}}{\partial\theta} &=&  
       \frac{\partial x_{n|n-1}}{\partial\theta} 
       \frac{\partial K_n}{\partial\theta} \varepsilon_n
       + K_n\frac{\partial \varepsilon_n}{\partial\theta}   \nonumber \\
\frac{\partial V_{n|n}}{\partial\theta} &=& 
         (I- K_nH)\frac{\partial V_{n|n-1}}{\partial\theta}
       - \frac{\partial K_n}{\partial\theta} H V_{n|n-1}. \nonumber
\end{eqnarray}

For Whard (whole sale hardware) data (Kitagawa (2020)), $N=155$, 
the standard seasonal adjustment model with $m_1=2$, $m_2=1$ is estimated
using the initial estimates of parameters, 
$\theta = (\log\tau_1^2, \log\tau_2^2, \log\sigma^2) =(-12.20607265, -13.81551056, -0.69314718)^T$.
The log-likelihood of the model with these initial parameters is $\ell (\theta )=109.34479$
and the Gradient obtained by the numerical difference function FUNCND and the proposed
method are shown in the Table \ref{Tab_Case1}.
In the numerical differentiation, $C=10^{-3}$is used.
It can be seen that the numerical differentiation coincides with the analytic
derivative up to 5th digit.

\begin{table}[h]
\begin{center}
\caption{Comparison of numerical diffference and gradient}\label{Tab_Case1}

\vspace{2mm}
\begin{tabular}{c|cc}
        & Numerical Difference &  Gradient \\
\hline
  $\frac{\partial \ell(\theta )}{\partial \tau_1^2}$ & 1.07694445 & 1.07694205 \\[2mm]
  $\frac{\partial \ell(\theta )}{\partial \tau_2^2}$ & 0.00091259 & 0.00091256 \\[2mm]
  $\frac{\partial \ell(\theta )}{\partial \sigma^2}$ & 70.91720451& 70.91720448 \\
\hline
\end{tabular}
\end{center}
\end{table}

\subsection{Seasonal adjustment model with stationary AR component}

Consider a seasonal adjustment model with statinary AR component
\begin{eqnarray}
y_n = T_n + S_n + p_n +w_n,
\end{eqnarray}
where $T_n$ and $S_n$ are the trend component and the seasonal component
introduced in the previous subsection and $p_n$ is an AR component with order $m_3$ defined my
\begin{eqnarray}
 p_n = \sum_{j=1}^{m_3} a_j p_{n-j} + v_n^{(t)}. 
\end{eqnarray}
Here $v_n^{(t)}$is a Gaussian white noise with variance $\tau_3^2$.
The model contains $4+m_3$ parameters and the parameter vector is given by
$\theta =(\theta_1 ,\ldots ,\theta_{m_3+4})\equiv (\tau_1^2,\tau_2^2,\tau_3^2,\sigma^2, a_1,\cdots ,a_{m_3})^T$.

The matrices $F$, $G$, $H$, $Q$ and $R$ are defined by
{\setlength{\arraycolsep}{1mm}
\begin{eqnarray}
x_n = \left[ \begin{array}{c} 
             T_n     \\
             T_{n-1} \\
             S_n     \\
             S_{n-1} \\
             \vdots  \\ 
             S_{n-p+2} \\
             p_{n-1}  \\
             p_{n-2}  \\
             \vdots   \\
             p_{n-m_3} \end{array}\right],\quad
F &=& \left[ \begin{array}{ccccccccccc} 
             2 &-1 &   &   &   &   \\
             1 & 1 &   &   &   &   \\
               &   &-1 &-1 &\cdots&-1\\
               &   & 1 &   &   &   \\
               &   &   &\ddots&&   \\ 
               &   &   &   & 1 &   \\
               &   &   &   &   &   &a_1 & a_2 &\cdots & a_{m_3} \\
               &   &   &   &   &   & 1  &     &       &    \\
               &   &   &   &   &   &    & \ddots &    &    \\
               &   &   &   &   &   &    &        & 1  &
         \end{array}\right],\quad
G = \left[ \begin{array}{ccc} 
             1 & 0 & 0  \\
             0 & 0 & 0 \\
             0 & 1 & 0 \\
             0 & 0 & 0 \\
          \vdots&\vdots&\vdots\\ 
             0 & 0 & 0 \\
             0 & 0 & 1 \\
             0 & 0 & 0 \\
          \vdots&\vdots\\ 
             0 & 0 & 0          
 \end{array}\right] \\
H &=& [\begin{array}{cccccccccc} 1&0&1&0&\cdots &0&1&0&\cdots &0 \end{array}] \nonumber \\
Q &=& \left[ \begin{array}{ccc} 
             \tau_1^2 & 0        & 0  \\
               0      & \tau_2^2 & 0  \\
               0      & 0  & \tau_3^2\end{array}\right], \quad
R = \sigma^2.
\end{eqnarray}  
}

In this case, all of $F$, $G$, $H$, $Q$ and $R$ are time-invariant and
do not depend on time $n$.
The relation between the parameter $\theta_j$ and the variances and AR coefficients 
are as follows.
\begin{eqnarray}
&& \tau_j^2 = e^{\theta_j}, (j=1,\ldots ,3), \quad
 \sigma^2 = e^{\theta_4}, \quad \beta_j = C\frac{e^{\theta_{j+4}}-1}{e^{\theta_{j+4}}+1}\\
&& \{a_j^{(m)} = a_j^{(m-1)} - \beta_m a_{m-j}^{(m-1)}, j=1,\ldots ,m \}, \quad \mbox{for }m=1,\ldots ,m_3
\label{Eq_Levinson}
\end{eqnarray}
Note that the equation (\ref{Eq_Levinson}) is the relation between the AR coefficients 
of order $m-1$ and those  of the order $m$ used in the Levinson's algorithm (Kitagawa (2020)).

In this case, 
\begin{eqnarray}
  \frac{\partial Q}{\partial\theta_1} &=&\left[ \begin{array}{ccc} 
             \tau_1^2 & 0 & 0  \\
               0      & 0 & 0  \\
               0      & 0 & 0  \end{array}\right], \quad
  \frac{\partial Q}{\partial\theta_2} =\left[ \begin{array}{ccc} 
               0 & 0  & 0 \\
               0 & \tau_2^2 & 0 \\
               0 & 0  & 0\end{array}\right], \quad
   \frac{\partial Q}{\partial\theta_3} =\left[ \begin{array}{ccc} 
               0 & 0  & 0  \\
               0 & 0  & 0  \\
               0 & 0  & \tau_3^2\end{array}\right], \nonumber\\
\frac{\partial Q}{\partial\theta_4} &=&\left[ \begin{array}{ccc} 
               0 & 0 & 0 \\
               0 & 0 & 0 \\
               0 & 0 & 0 \\
 \end{array}\right], \quad 
  \frac{\partial R}{\partial\theta_1} = 0, \quad
  \frac{\partial R}{\partial\theta_2} = 0, \quad
  \frac{\partial R}{\partial\theta_3} = 0, \quad
  \frac{\partial R}{\partial\theta_4} = \sigma^2, \\
\left(\frac{\partial F}{\partial \theta_k}\right)_{pq} &=&
   \left\{ \begin{array}{cl} \displaystyle
             \frac{\partial a_p}{\partial \theta_q} & \mbox{if }k>4, p=4, q=5,\ldots ,m_3+4 \\
             0                                      & \mbox{otherwise} \end{array}
   \right.
\end{eqnarray}
where $\displaystyle\left(\frac{\partial F}{\partial \theta_k}\right)_{pq}$ denotes the $(p,q)$ component
of the matrix $\displaystyle\left(\frac{\partial F}{\partial \theta_k}\right)$ , and
$\displaystyle\frac{\partial a_i^{(m)}}{\partial\theta_j} $ is obtained by
\begin{eqnarray}
\frac{\partial a_i^{(m)}}{\partial\theta_j} 
       =  \frac{\partial a_i^{(m)}}{\partial\beta_j} 
          \frac{\partial\beta_j}{\partial \theta_j}
       = 2C \frac{e^{\theta_j}}{(e^{\theta_j}+1)^2}\frac{\partial a_i^{(m)}}{\partial\beta_j} 
          ,\quad j=1,\ldots ,m
\end{eqnarray}
and 
\begin{eqnarray}
 \frac{\partial\beta_k}{\partial \theta_j}
   &=& \left\{  \begin{array}{ll}
       \displaystyle 2C \frac{e^{\theta_j}}{(e^{\theta_j}+1)^2}  & \mbox{for }k=j \\
       \displaystyle 0                                          & \mbox{for }k\neq j
       \end{array}  \right.
\\
\frac{\partial a_i^{(m)}}{\partial\beta_k} 
    &=& \left\{  \begin{array}{ll}
          0         & \mbox{for }i=m \mbox{ and } k<m \\[3mm]
          1         & \mbox{for }i=m=k \\[2mm]
        \displaystyle \frac{\partial a_i^{(m-1)}}{\partial\beta_k}
            - \beta_m \frac{\partial a_{m-i}^{(m-1)}}{\partial\beta_k}
         & \mbox{for }i<m\mbox{ and }k <m \\[4mm]
           -  a_{m-i}^{(m-1)} 
         & \mbox{for }i<m\mbox{ and }k=m. \end{array}
        \right.
\end{eqnarray}

Table \ref{Tab_Case2} shows the gradients obtained by the numerical difference and the 
proposed method. 
The initial estimates of the parameters are 
$\theta = (-12.20607265,-13.81551056, -9.72116600,$ $-0.69314718, 2.92316158, -1.20485737)$
and the log-likelihood of the model is
$\ell (\theta )$ = 109.39234337.
In this case as well, the numerical differentiation matches the analytic derivative up to the fifth digit.

\begin{table}[h]
\begin{center}
\caption{Comparison of numerical diffference and gradient}\label{Tab_Case2}

\vspace{2mm}
\begin{tabular}{c|cc}
        & Numerical Difference &  Gradient \\
\hline
  $\frac{\partial \ell(\theta )}{\partial \tau_1^2}$ & 1.07570844 & 1.07570605 \\[2mm]
  $\frac{\partial \ell(\theta )}{\partial \tau_2^2}$ & 0.00091252 & 0.00091249 \\[2mm]
  $\frac{\partial \ell(\theta )}{\partial \tau_3^2}$ & 0.04739855 & 0.04739781 \\[2mm]
  $\frac{\partial \ell(\theta )}{\partial \sigma^2}$ & 70.87177866& 70.87177864 \\[2mm]
  $\frac{\partial \ell(\theta )}{\partial a_1}$ &  0.03112269 & 0.03112271 \\[2mm]
  $\frac{\partial \ell(\theta )}{\partial a_2}$ & -0.02850531 & -0.02850530 \\
\hline
\end{tabular}
\end{center}
\end{table}

\subsection{ARMA Model}
Consider a stationary ARMA model\index{ARMA model} (autoregressive moving average model)\index{autoregressive moving average model} of order $(m, \ell)$ (Box and Jenkins (1970), Brockwell and Davis (1981))%
\begin{equation}
 y_n = \sum_{j=1}^{m}a_{j}y_{n-j}+v_{n}-\sum_{j=1}^{\ell}b_{j}v_{n-j},
\end{equation}
where $v_n$ is a Gaussian white noise with mean zero and variance $\sigma^2$.
Here, a new variable $\tilde{y}_{n+i|n-1}$ is defined as
\begin{equation}
 \tilde{y}_{n+i|n-1} = \sum_{j=i+1}^{m}a_{j}y_{n+i-j}-\sum_{j=i}^{\ell}b_{j}v_{n+i-j}, 
\end{equation}

\noindent
which is a part of $y_{n+i}$ that can be directly computable from the observations until time $n-1$, $y_{n-1}$, $y_{n-2}$, $\cdots$, and the noise inputs until time $n$, $v_n$, $v_{n-1}$, $\cdots$.

By setting  $k=\max (m,\ell+1)$ and defining the $k$-dimensional state vector $x_n$ as
\begin{equation}
 x_n = ( y_n,\tilde{y}_{n+1|n-1},\cdots ,\tilde{y}_{n+k-1|n-1} )^T,
\end{equation}
\noindent
the ARMA model can be expressed in the form of a state-space model (Kitagawa (2020)): 
\begin{eqnarray} 
 x_n &=& Fx_{n-1}+Gv_n \nonumber \\
 y_n &=& Hx_n.
\end{eqnarray}
Here $k= \max (m,\ell +1)$ and the $k$ $\times$ $k$ matrix $F$ and the $k$-dimensional vectors $G$ and $H$ are defined as
%
%
\begin{eqnarray}
 F &=& \left[ \begin{array}{cccc}
 a_1 & 1 & & \\
 a_2 & & \ddots& \\
 \vdots & & & 1 \\
 a_k & & & \end{array}\right], \hspace{5mm}
 G = \left[ \begin{array}{c} 1 \\[1mm] -b_{1} \\ \vdots \\[1mm] -b_{k-1} 
 \end{array}\right] \\
 H &=&\;
[ \begin{array}{cccc} \,1 & 0 & \;\cdots\; & 0\,
 \end{array}], \nonumber
\end{eqnarray}
\noindent
respectively, where $a_{i} =0$ for $i>m$ and $b_{i}$ = 0 for $i>\ell$.

The ARMA model of order $(m, \ell )$ has $m+\ell + 1$ unknown parameters 
$\sigma^2 , a_1,\ldots ,a_m, b_1,\ldots ,b_\ell $.
However, the maximum likelihood estimate of the innovation variance is
obtained by
\begin{eqnarray}
 \hat\sigma^2 =\frac{1}{N}\sum_{n=1}^N \frac{\varepsilon_n^2}{r_n},
\end{eqnarray}
and the coefficents $a_i$ and $b_j$ can be estimated independent on
the varaince.
Therefore, hereafter define the paramter vector as
$\theta = (a_1,\ldots ,a_m, b_1,\ldots ,b_\ell )^T$.
Then the log-likelihood of the ARMA model is given by
\begin{eqnarray}
 \ell (\theta ) = 
    - \frac{1}{2} \biggl\{ N \log 2\pi +  N \log \hat\sigma^2
    + \sum_{n=1}^N \log r_{n} + N \biggr\}. \label{Eq_log-lk_ARMA}
\end{eqnarray}
%

\subsubsection{Gradient filter for ARMA model}
For the state-space representation of the ARMA model, the derivative of the matrices  $F$, $G$, and $Q$ are given by
\begin{eqnarray}
&&\frac{\partial F_{ij}}{\partial\theta_p} 
     = \left\{ \begin{array}{cl}  \displaystyle
           1   &\quad \mbox{if } 2\leq i \leq m, j=1, p\leq m\\[2mm]
           0   &\quad \mbox{otherwise}
       \end{array} \right.  \\
&&\frac{\partial G_{i}}{\partial\theta_p} 
     = \left\{ \begin{array}{cl}  \displaystyle
          -1  &\quad \mbox{if } i \leq \ell , m+1 \leq p \leq m+\ell \\[2mm]
           0  &\quad \mbox{otherwise}
       \end{array} \right.  . \\
&& \frac{\partial Q}{\partial\theta_p} =  0 ,\qquad p=1,\ldots ,m+\ell 
\end{eqnarray}

From (\ref{Eq_log-lk_ARMA}), the gradient of the log-likelihood of the ARMA model is obtained by
\begin{eqnarray}
\frac{\partial\ell (\theta )}{\partial \theta} 
&=&     - \frac{1}{2}\sum_{n=1}^N \frac{1}{r_n}\frac{\partial r_n}{\partial\theta}
     - \frac{1}{\hat\sigma^2} 
          \sum_{n=1}^N \frac{\varepsilon_n}{r_n}  \frac{\partial \varepsilon_n}{\partial\theta}
       +\frac{1}{2\hat\sigma^2} \sum_{n=1}^N \frac{\varepsilon_n^2}{r_n^2}\frac{\partial r_n}{\partial\theta},
\end{eqnarray}
where the derivatives of the one-step-ahead predition error $\varepsilon_n$ and the
one-step-ahead prediction error variance $r_n$ are obtained by
\begin{eqnarray}
\frac{\partial \varepsilon_n}{\partial\theta} &=& -H \frac{\partial x_{n|n-1}}{\partial\theta}
  \nonumber \\
\frac{\partial r_n}{\partial\theta} &=& H \frac{\partial V_{n|n-1}}{\partial\theta}H^T .
\end{eqnarray}
Here $\displaystyle \frac{\partial x_{n|n-1}}{\partial\theta}$ and 
$\displaystyle \frac{\partial V_{n|n-1}}{\partial\theta}$ can be evaluated
by the following Kalman filter like recursive algorithm

\quad [One-step-ahead-prediction] \\
\begin{eqnarray}
\frac{\partial x_{n|n-1}}{\partial\theta} &=& F \frac{\partial x_{n-1|n-1}}{\partial\theta}
   + \frac{\partial F}{\partial\theta} x_{n-1|n-1} .
\nonumber \\
\frac{\partial V_{n|n-1}}{\partial\theta} &=& F \frac{\partial V_{n-1|n-1}}{\partial\theta}F^T 
           + \frac{\partial F}{\partial\theta}V_{n-1|n-1}F^T 
           + F V_{n-1|n-1} \frac{\partial F}{\partial\theta}^T \nonumber \\
         &&   + \frac{\partial G}{\partial\theta}Q G^T 
           + G Q \frac{\partial G}{\partial\theta}^T .  \label{Eq_gradient-filter-P}
\end{eqnarray}

\quad [Filter] \\
\begin{eqnarray}
\frac{\partial K_n}{\partial\theta} 
        &=& \frac{\partial V_{n|n-1}}{\partial\theta}H^T r_n^{-1} 
          - V_{n|n-1}H^Tr_n^{-2} 
          \frac{\partial r_n}{\partial\theta}    \nonumber \\
\frac{\partial x_{n|n}}{\partial\theta} &=&  
        \frac{\partial x_{n|n-1}}{\partial\theta}
       + K_n \frac{\partial \varepsilon_n}{\partial\theta}
       + \frac{\partial K_n}{\partial\theta} \varepsilon_n   \nonumber \\
\frac{\partial V_{n|n}}{\partial\theta} &=& 
         \frac{\partial V_{n|n-1}}{\partial\theta}
       - \frac{\partial K_n}{\partial\theta} H V_{n|n-1}
       - K_n H \frac{\partial V_{n|n-1}}{\partial\theta} \label{Eq_gradient-filter-F}
\end{eqnarray}

To apply the above recursive agorithm, we need the initial values,
$\displaystyle \frac{\partial V_{ij}}{\partial a_p}$ and 
$\displaystyle \frac{\partial V_{ij}}{\partial b_r}$ ,
which can be obtained from the initial variance covaiance matrix of the
state-space representaion of  the ARMA model (Kitagawa (2020))
\begin{eqnarray}
  V_{11} &=&  C_0 \nonumber \\
  V_{1i} &=& \sum_{j=i}^m a_jC_{j+1-i}\; - 
             \sum_{j=i-1}^\ell b_jg_{j+1-i} \nonumber \\
  V_{ij} &=&  \sum_{p=i}^m\sum_{q=j}^m a_p a_q C_{q-j-p+i}\;
          -  \sum_{p=i}^m\sum_{q=j-1}^\ell a_p b_q g_{q-j-p+i}             \nonumber \\              
        & &- \sum_{p=i-1}^\ell\sum_{q=j}^m b_p a_q g_{p-i-q+j}  
           +  \sum_{p=i-1}^\ell b_p b_{p+j-i} \sigma^2 , 
\end{eqnarray}
where the autocovariace function $C_k, k=0,1,\ldots ,k$ and the impulse response function
$g_k$ are obtained by

[Impulse response function]
\begin{eqnarray}
  g_0 &=& 1 \nonumber \\
  g_i &=& \sum_{j=1}^i a_j g_{i-j} - b_i, \hspace{8mm}i=1,2,\cdots \label{eq_ARMA_impulse}
\end{eqnarray}

[Covariance function]
\begin{eqnarray}
  C_0 &=& \sum_{i=1}^m a_i C_i + \sigma^2 \biggl( 1 -  \sum_{i=1}^{\ell} b_i g_i \biggr) \label{Eq_ARMA_autocovariance_0}\\
  C_k &=& \sum_{i=1}^m a_i C_{k-i} -\sigma^2 \sum_{i=k}^{\ell} b_i g_{i-k},
\hspace{8mm}k=1,2,\cdots  \label{Eq_ARMA_autocovariance}
\end{eqnarray}

\subsubsection{Initial condition for the recursive computation}
To apply this recursive agorithm shown in (\ref{Eq_gradient-filter-P}) and (\ref{Eq_gradient-filter-F}), we need the initial values,
$\displaystyle \frac{\partial V_{ij}}{\partial a_p}$ and 
$\displaystyle \frac{\partial V_{ij}}{\partial b_r}$ 
which are obtained as
\begin{eqnarray}
  \frac{\partial V_{11}}{\partial a_p} &=&   \frac{\partial C_0}{\partial a_p},\quad
  \frac{\partial V_{11}}{\partial b_p} =   \frac{\partial C_0}{\partial b_p}, \nonumber \\
%
\frac{\partial V_{1i}}{\partial a_p} &=& \sum_{j=i}^m a_j \frac{\partial C_{j+1-i}}{\partial a_p} + C_{p+1-i} - \sum_{j=i-1}^\ell b_j \frac{\partial g_{j+1-i}}{\partial a_p} \nonumber \\
\frac{\partial V_{1i}}{\partial b_p} &=& \sum_{j=i}^m a_j \frac{\partial C_{j+1-i}}{\partial b_p} - \sum_{j=i-1}^\ell b_j \frac{\partial g_{j+1-i}}{\partial b_p} - g_{p+1-i} \nonumber \\
%
%
%
 \frac{\partial V_{ij}}{\partial a_r}
     &=&   \sum_{p=i}^m a_p C_{r-j-p+i}
        +  \sum_{q=j}^m a_q C_{q-j-r+i}     
        +  \sum_{p=i}^m\sum_{q=j}^m a_p a_q \frac{\partial C_{q-j-p+i}}{\partial a_r}      \nonumber \\
     && -  \sum_{q=j-1}^\ell b_q g_{q-j-r+i}                   
        -  \sum_{p=i}^m\sum_{q=j-1}^\ell a_p b_q \frac{\partial g_{q-j-p+i}}{\partial a_r}   \nonumber \\              
     &&- \sum_{p=i-1}^\ell b_p g_{p-i-r+j}  
       - \sum_{p=i-1}^\ell\sum_{q=j}^m b_p a_q \frac{\partial g_{p-i-q+j}}{\partial a_r}\\
 \frac{\partial V_{ij}}{\partial b_r}
    &=&  \sum_{p=i}^m\sum_{q=j}^m a_p a_q \frac{\partial C_{q-j-p+i}}{\partial b_r}
          -  \sum_{p=i}^m a_p g_{r-j-p+i}  
          -  \sum_{p=i}^m\sum_{q=j-1}^\ell a_p b_q \frac{\partial g_{q-j-p+i}}{\partial b_r}             \nonumber \\ 
    & &- \sum_{q=j}^m a_q g_{r-i-q+j}  
       - \sum_{p=i-1}^\ell\sum_{q=j}^m b_p a_q \frac{\partial g_{p-i-q+j}}{\partial b_r}  \nonumber \\
    & & +  \sum_{p=i-1}^{\ell} b_{r+j-i} \sigma^2  
        +  \sum_{p=i-1}^{\ell} b_{r-j+i} \sigma^2 .
\end{eqnarray}

Here, from the definition of the impulse responce function (\ref{eq_ARMA_impulse}) and the 
autocovariance function (\ref{Eq_ARMA_autocovariance_0}) and (\ref{Eq_ARMA_autocovariance}), their derivatives are obtained as
follows:
\begin{eqnarray}
  \frac{\partial g_0}{\partial\theta_j} &=& 0 ,\qquad j=1,\ldots m+\ell\nonumber \\
  \frac{\partial g_i}{\partial a_p} &=& \sum_{j=1}^i a_j \frac{\partial g_{i-j}}{\partial a_p} + g_{i-p}, \hspace{8mm}i=1,2,\cdots \label{arma-1-41} \\
 \frac{\partial g_i}{\partial b_p} &=& \left\{ \begin{array}{ll} \displaystyle 
      \sum_{j=1}^i a_j \frac{\partial g_{i-j}}{\partial b_p} , &i=1,2,\cdots  \\[2mm]
       \displaystyle
      \sum_{j=1}^i a_j \frac{\partial g_{i-j}}{\partial b_p}  - 1 , & i=1,2,\cdots ,\quad p=i
  \end{array}
  \right. \nonumber
\end{eqnarray}

%
\begin{eqnarray}
%
  \frac{\partial C_0}{\partial a_p} &=& C_p + \sum_{i=1}^m a_i \frac{\partial C_i}{\partial a_p}  
  - \sigma^2 \sum_{i=1}^{\ell} b_i \frac{\partial g_i}{\partial a_p} \label{Eq_dC0da}\\
  \frac{\partial C_0}{\partial b_p} &=& \sum_{i=1}^m a_i \frac{\partial C_i}{\partial b_p}    - \sigma^2 \biggl( g_p + \sum_{i=1}^{\ell} b_i \frac{\partial g_i}{\partial b_p} \biggr) \label{Eq_dC0db}\\
\frac{\partial C_k}{\partial a_p} &=& C_{k-p} + \sum_{i=1}^m a_i \frac{\partial C_{k-i}}{\partial a_p} -\sigma^2 \sum_{i=k}^{\ell} b_i \frac{\partial g_{i-k}}{\partial a_p},
\hspace{8mm}k=1,2,\cdots  \label{Eq_dCda} \\
\frac{\partial C_k}{\partial b_p} &=& \sum_{i=1}^m a_i \frac{\partial C_{k-i}}{\partial b_p} -\sigma^2 \biggl(g_{p-k} + \sum_{i=k}^{\ell} b_i \frac{\partial g_{i-k}}{\partial b_p}\biggr),
\hspace{8mm}k=1,2,\cdots  \label{Eq_dCdb} 
\end{eqnarray}

Note that the equations (\ref{Eq_dC0da}) -- (\ref{Eq_dCdb}) are expressed in the following form 
and can be solved in the same way as the equations (\ref{Eq_ARMA_autocovariance_0}) and (\ref{Eq_ARMA_autocovariance}).
%
\setlength{\arraycolsep}{1mm}
%
%
\begin{eqnarray}
\renewcommand{\arraystretch}{1.5}
\left[  \begin{array}{c} \frac{\partial C_0}{\partial a_p} \\
    \frac{\partial C_1}{\partial a_p} \\
    \vdots \\
    \frac{\partial C_k}{\partial a_p} 
    \end{array}\right] 
= \left[  \begin{array}{cccc} 
    0   & a_1 & \cdots & a_k \\
    a_1 & a_2 & \cdots & a_{k-1}\\
    \vdots & \vdots &  &\vdots \\
    a_k & a_{k-1}&\cdots& 0  \end{array}\right] 
  \left[  \begin{array}{c} \frac{\partial C_0}{\partial a_p} \\
    \frac{\partial C_1}{\partial a_p} \\
    \vdots \\
    \frac{\partial C_k}{\partial a_p}
    \end{array}\right] 
+ \left[  \begin{array}{c} 
    C_p - \sigma^2 \sum_{i=p}^{\ell} b_i \frac{\partial g_{i-p}}{\partial a_p}  \\
    C_{p-1}- \sigma^2 \sum_{i=p+1}^{\ell} b_i \frac{\partial g_{i-p-1}}{\partial a_p} \\
    \vdots \\
    C_{p-k} - \sigma^2 b_\ell \frac{\partial g_{\ell -i}}{\partial a_p}
    \end{array}\right]
\end{eqnarray}
%
%
%
\begin{eqnarray}
\renewcommand{\arraystretch}{1.5}
\left[  \begin{array}{c} \frac{\partial C_0}{\partial b_p} \\
    \frac{\partial C_1}{\partial b_p} \\
    \vdots \\
    \frac{\partial C_k}{\partial b_p} 
    \end{array}\right] 
= \left[  \begin{array}{cccc} 
    0   & a_1 & \cdots & a_k \\
    a_1 & a_2 & \cdots & a_{k-1}\\
    \vdots & \vdots &  &\vdots \\
    a_k & a_{k-1}&\cdots& 0  \end{array}\right] 
  \left[  \begin{array}{c} \frac{\partial C_0}{\partial b_p} \\
    \frac{\partial C_1}{\partial b_p} \\
    \vdots \\
    \frac{\partial C_k}{\partial b_p}
    \end{array}\right] 
+ \left[  \begin{array}{c} 
  - \sigma^2 \biggl( g_p + \sum_{i=1}^{\ell} b_i \frac{\partial g_i}{\partial b_p} \biggr)  \\
  - \sigma^2 \biggl( g_{p-1} + \sum_{i=1}^{\ell} b_i \frac{\partial g_{i-1}}{\partial b_p} \biggr)  \\
    \vdots \\
     - \sigma^2 \biggl( g_{p-k} + \sum_{i=1}^{\ell} b_i \frac{\partial g_{i-k}}{\partial b_p} \biggr)
    \end{array}\right] .
\end{eqnarray}

\subsubsection{Effect of transformation of parameters}
In actual parameter estimation, however, to satisfy the stationarity and invertibility conditions, we usually apply the following transformations of the parameters. 

For the condition of stationarity\index{stationarity} for the AR coefficients $a_1, \cdots, a_m$, associated partial autocorrelation coefficients $\beta_1, \cdots, \beta_m$ should satisfy $-1 < \beta_i < 1$ for all $i=1, \cdots, m$. It can be seen that this condition is guaranteed, if the transformed coefficients $\alpha_i$ defined by 
\begin{equation}
 \alpha_i = \log \biggl( \frac{1+\beta_i}{1-\beta_i} \biggr),
\end{equation}
satisfy $-\infty <\alpha_i <\infty$ for all $i=1,\cdots,m$.

Conversely, if $\beta_i$ is defined by 
\begin{equation}
 \beta_i = \frac{e^{\alpha_i}-1}{e^{\alpha_i}+1}, \label{eq_ARMA_est_PARCOR}
\end{equation}
\noindent
for arbitrary $(\alpha_1,\cdots,\alpha_m) ^T \in {\rm R} ^m$, then it can been seen that $|\beta_i|<1$ is always satisfied and the corresponding AR coefficients satisfy the stationarity condition.

Similarly, to guarantee the invertibility\index{invertibility} condition of the MA coefficents for any $(\delta_1,\cdots,\delta_\ell ) ^T \in {\rm R} ^\ell$, let $\gamma_i$ be defined as
\begin{equation}
 \gamma_i = \frac{e^{\delta_i}-1}{e^{\delta_i}+1},
\end{equation}
and formally obtain the corresponding MA coefficients $b_1, \cdots, b_\ell$ by considering $d_1, \cdots, d_\ell$ to be the PARCOR's.

Then for arbitrary $\theta^{\prime\prime} = (\alpha_1, \cdots, \alpha_m, \gamma_1, \cdots, \gamma_\ell)^T \in {\rm R}^{m+\ell}$, the corresponding ARMA model will always satisfy the stationarity and invertibility conditions.
It is noted that if the coefficient needs to satisfy the
condition that $|\beta_i|<C$ for some $0 < C <1$, we define 
\begin{eqnarray}
\beta_i = \frac{e^{\alpha_i}-1}{e^{\alpha_i}+1} C,
\end{eqnarray}
instead of the equation (\ref{eq_ARMA_est_PARCOR}).

For these transformations, the gradient of the log-likelihood is modified as follows.
\begin{eqnarray}
\frac{\partial\ell (\theta)}{\partial\theta_j} = \left\{ 
     \begin{array}{ll} \displaystyle \sum_{i=1}^m
         \frac{\partial\ell (\theta)}{\partial a_i}\frac{\partial a_i}{\partial \theta_j} 
          & {}\quad\mbox{for } j=1,\ldots ,m \\[4mm]
          \displaystyle \sum_{i=1}^{-\ell}
         \frac{\partial\ell (\theta)}{\partial b_{i}}\frac{\partial b_{i}}{\partial \theta_j} 
          & {}\quad\mbox{for } j=m+1,\ldots ,\ell \\
     \end{array}
\right.
\end{eqnarray}
where $ \displaystyle\frac{\partial a_i}{\partial\theta_{j}}$ and $ \displaystyle\frac{\partial b_{i}}{\partial\theta_{j}}$
are obtained by
\begin{eqnarray}
\frac{\partial a_j^{(m)}}{\partial\theta_j} 
       &=&  \frac{\partial a_i^{(m)}}{\partial\beta_j} 
          \frac{\partial\beta_j}{\partial \theta_j}
       =  \frac{2Ce^{\theta_j}}{(e^{\theta_j}+1)^2}\frac{\partial a_i^{(m)}}{\partial\beta_j} 
          ,\quad j=1,\ldots ,m \\
\frac{\partial b_i^{(m)}}{\partial\theta_{m+j}} 
       &=&  \frac{\partial b_i^{(m)}}{\partial\gamma_j} 
          \frac{\partial\gamma_j}{\partial \theta_{m+j}}
       =  \frac{2Ce^{\theta_{m+j}}}{(e^{\theta_{m+j}}+1)^2}\frac{\partial b_i^{(m)}}{\partial\gamma_j} 
          ,\quad j=1,\ldots ,\ell ,
\end{eqnarray}
and 
$\displaystyle\frac{\partial a_i^{(m)}}{\partial\beta_j}$ and 
$\displaystyle\frac{\partial b_i^{(m)}}{\partial\gamma_j}$
are given by

\begin{eqnarray}
\frac{\partial a_i^{(m)}}{\partial\beta_k} 
    &=& \left\{  \begin{array}{ll}
          0         & \mbox{for }i=m \mbox{ and } k<m \\[2mm]
          1         & \mbox{for }i=m=k \\[2mm]
       \displaystyle
          \frac{\partial a_i^{(m-1)}}{\partial\beta_k}
            - \beta_m \frac{\partial a_{m-i}^{(m-1)}}{\partial\beta_k}
         & \mbox{for }i<m\mbox{ and }k <m \\[5mm]
           -  a_{m-i}^{(m-1)}
         & \mbox{for }i<m\mbox{ and }k=m. \end{array}
        \right.  \\
\frac{\partial b_i^{(\ell )}}{\partial\gamma_k} 
    &=& \left\{  \begin{array}{ll} 
          0         & \mbox{for }i=\ell \mbox{ and } k<\ell \\[2mm]
          1         & \mbox{for }i=\ell =k \\[2mm]
        \displaystyle
          \frac{\partial b_i^{(\ell -1)}}{\partial\gamma_k}
            - \gamma_\ell \frac{\partial b_{\ell -i}^{(\ell -1)}}{\partial\gamma_k}
         & \mbox{for }i<\ell\mbox{ and }k <\ell \\[5mm]
           -  b_{\ell -i}^{(\ell -1)} 
         & \mbox{for }i<\ell\mbox{ and }k=\ell . \end{array}
        \right. 
\end{eqnarray}


\subsubsection{ARMA(2,1) and ARMA(5,3)}

As numerical  examples, we consider two ARMA models for the Hakusan yaw rate
data (Kitagawa (2020)).
The first example is an ARMA(2,1) model.
The initial estimates of the AR and MA coefficents are $a_1=1.3$, $a_2=-0.6$, $b_1=0.2$.
The log-likelihood of the ARMA model with these initial parameters are $-16.3976$.
Table \ref{Tab_ARMA(2,1)} compare the gradients of the log-likelihood 
computed by the numerical difference
and the proposed gradient filter algorithm.
The gradients coincides until the fifth digit.
By both algorithm, the maximum likelihood estimates of the model are
$a_1=1.4103$, $a_2=-0.6846$, $b_1=0.3396$, $\sigma^2 = 0.06663$ and the
maximum log-likelihood $\ell (\hat\theta )=-15.7187$, AIC = 39.4373.

\begin{table}[h]
\begin{center}
\caption{Comparison of numerical diffference and gradient for ARMA(2,1) model.}
\label{Tab_ARMA(2,1)}

\vspace{2mm}
\begin{tabular}{c|cc}
        & Numerical Difference &  Gradient \\
\hline
  $\frac{\partial \ell(\theta )}{\partial \theta_1}$ & $-0.7848665$ & $-0.7848652$ \\[2mm]
  $\frac{\partial \ell(\theta )}{\partial \theta_2}$ & $\;\:\: 1.6988569$ & $\;\:\: 1.6988567$ \\[2mm]
  $\frac{\partial \ell(\theta )}{\partial \theta_3}$ & $-1.6783890$ & $-1.6783890$ \\
\hline
\end{tabular}
\end{center}
\end{table}

The second example is the ARMA(5,3) model for the same data set.
Initial estimates of the parameters are
$a_1 = 2.5$, $a_2 =-3.0$, $a_3 = 2.1$, $a_4 =-1.0$, $a_5 = 0.3$,
$b_1= 2.1$, $b_2=-1.7$, $b_3= 0.5$ and the log-likelihood of the
model with these parameters is $\ell = -156.5930$.
Table \ref{Tab_ARMA(5,3)} compare the gradients of the log-likelihood 
computed by the numerical difference
and the proposed algorithm.
The gradients coincides at least until the sixth digit.
By both algorithm, the maximum likelihood estimates of the model are
$a_1=3.0705$, $a_2=-4.0905$, $a_3 = 2.9810$, $a_4=-1.27978$, $a_5= 0.3035$, 
$b_1=2.982$, $b_2=-1.6797$, $b_3= 0.5023$, $\sigma^2 = 0.05743$ and the
maximum log-likelihood $\ell (\hat\theta )= 0.8624$, AIC = 16.2753.
The AIC values indicate the ARMA(5,3) is better than the ARMA(2,1) model.

\begin{table}[h]
\begin{center}
\caption{Comparison of numerical diffference and gradient for ARMA(5,3) model.}
\label{Tab_ARMA(5,3)}

\vspace{2mm}
\begin{tabular}{c|rr}
        & Numerical Difference &  Gradient$\hspace*{10mm}$ \\
\hline
  $\frac{\partial \ell(\theta )}{\partial \theta_1}$ & $-0.249927228\times 10^3$ &$-0.249927233\times 10^3$ \\[2mm]
  $\frac{\partial \ell(\theta )}{\partial \theta_2}$ & $ 0.910195611\times 10^1$ &$ 0.910195568\times 10^1$ \\[2mm]
  $\frac{\partial \ell(\theta )}{\partial \theta_3}$ & $-0.342739937\times 10^2$ &$-0.342739934\times 10^2$ \\[2mm]
  $\frac{\partial \ell(\theta )}{\partial \theta_4}$ & $ 0.771826263\times 10^2$ &$ 0.771826264\times 10^2$ \\[2mm]
  $\frac{\partial \ell(\theta )}{\partial \theta_5}$ & $ 0.231448005\times 10^2$ &$ 0.231448006\times 10^2$ \\[2mm]
  $\frac{\partial \ell(\theta )}{\partial \theta_6}$ & $ 0.480755088\times 10^2$ &$ 0.480755057\times 10^2$ \\[2mm]
  $\frac{\partial \ell(\theta )}{\partial \theta_7}$ & $-0.850532732\times 10^2$ &$-0.850532748\times 10^2$ \\[2mm]
  $\frac{\partial \ell(\theta )}{\partial \theta_8}$ & $ 0.322328498\times 10^2$ &$ 0.322328498\times 10^2$ \\
\hline
\end{tabular}
\end{center}
\end{table}

\section{Summary}
The gradient and Hessian of the log-likelihood of linear state-space 
model are given. 
Details of the implementation of the algorithm for standard seasonal adjustment model,
seasonal adjustment model with stationary AR component and ARMA model are given.
For each implementation, comparison with a numerical difference method is shown.

\vspace{15mm}
\noindent{\Large\bf Aknowledgements}

This work was supported in part by JSPS KAKENHI Grant Number 18H03210.
The author is grateful to the project members, Prof. Kunitomo, Prof Nakano,
Prof. Kyo, Prof. Sato, Prof. Tanokura and Prof. Nagao for their 
stimulating discussions.

\end{document}